\documentclass[11pt,dvips]{article}
\usepackage{epsfig,times} 
\usepackage{picinpar}
\usepackage{wrapfig}
\usepackage{floatflt}
\makeatletter
\renewcommand{\section}{\@startsection{section}{1}{0in}
	{0.4\baselineskip}{0.1\baselineskip}{\Large\bf}}
\renewcommand{\subsection}{\@startsection{subsection}{2}{0in}
	{0.25\baselineskip}{-\baselineskip}{\large\bf}}
\renewcommand{\subsubsection}{\@startsection{subsubsection}{3}{0in}
	{0.1\baselineskip}{-\baselineskip}{\normalsize\bf}}
\makeatother
\begin{document}

\begin{flushleft}

HE. 5.3.02

\end{flushleft}

%

\begin{center}
{\LARGE \bf  Search for Magnetic Monopoles with MACRO}
\end{center}

\begin{center}
%
%
{\bf B.~C.~Choudhary$^1$ for the MACRO Collaboration\\}
{\it $^1$California Institute of Technology, Pasadena, CA 91125,
USA \\ }
\end{center}

\begin{center}
{\large \bf Abstract\\}
\end{center}
\vspace{-0.5ex}
%
%
Data from the complete MACRO detector 
have been used to search for magnetic monopoles (MMs)
of all expected velocities. Scintillator, 
 streamer tube (instrumented with specialized electronics)
and nuclear track detectors
have been used to search for signatures coming from MMs;  the scintillator and
nuclear track subdetectors were used also for searches for  
other rare particles (nuclearites, charged Q-balls).
 Based on no observation of such 
signals, we establish stringent flux limits,  for 
MMs  as slow as a few $10^{-5}c$.
%

\vspace{1ex}

\section{Introduction:}
\label{intro.sec}
Grand Unified Theories (GUTs) of the electroweak and strong interactions
predict the existence of massive ($ \sim 10^{17}$ GeV) magnetic monopoles,
 Preskill (1979).
 One of the primary aims of the MACRO detector at the Gran Sasso underground 
lab (in Italy, at an average depth of 3700 hg/cm$^2$)
is the search for such MMs
at a sensitivity level well below the
Parker bound ($10^{-15}$ cm$^{-2}$ s$^{-1}$ sr$^{-1}$), Turner et al. (1982),
for a
large range of velocities, $4 \cdot 10^{-5}<\beta<1$, $\beta=v/c$. \par
 MACRO uses three
different types of detectors: liquid scintillators, limited steamer tubes
and nuclear track detectors (CR39 and Lexan) arranged in a modular
structure of six ``supermodules" (SM's).
 Each SM is divided into a lower and an upper (``Attico") part.
The overall dimensions of the apparatus are $ 76.6 \times 12 \times 9$ 
m$^3$, Ahlen et al. (1993).
The response of the three types of detectors to slow and
fast particles was experimentally studied by Ahlen \& Tarl\'{e} (1983),
 Battistoni et al. (1988, 1997b) and
Cecchini et al.~(1996).
The 3 subdetectors ensure redundancy of information, cross-checks
and independent signatures for possible MM candidates. \par
The analyses reported here, obtained using the various subdetectors in a 
stand-alone and in a combined way, 
refer to direct detection of bare MMs of one
 unit Dirac
 charge ($g_D=137/2e$),  catalysis cross section $\sigma_{cat} <1$ mb
(we ignore monopole induced nucleon decay),
and  isotropic flux (we consider MMs with enough kinetic energy to
traverse the Earth); this last condition sets a $\beta$ dependent mass 
threshold ($\sim 10^{17}$ GeV for $\beta \sim 5 \cdot 10^{-5}$, and 
lower for faster MMs). \par
We also discuss the limits obtained for 
nuclearites and charged Q-balls.

\section{Searches for MMs with Individual Subdetectors:}
\subsection{Searches with Scintillators:} The searches with 
the liquid scintillator subdetector use different 
specialized triggers covering different velocity regions; the 
searches are grouped into searches for low velocity 
($10^{-4} < \beta < 10^{-3}$),
 medium velocity ($10^{-3} < \beta < 10^{-1}$) and high velocity
($\beta>0.1)$.

\subsubsection {Low Velocity Monopole Searches:}
Previous searches using  data collected with the Slow Monopole Trigger (SMT)
and  Waveform Digitizer (WFD)
were reported  in Ambrosio et al. (1997), see curves ``A", ``B'' in
Fig. \ref{fig:limiti}a. 
 A new 200 MHz WFD system was implemented which  improves by at 
least a factor of two the sensitivity
to very slow monopoles ($\beta\sim10^{-4}$) and by over a factor of five
the sensitivity to relativistic monopoles with 
respect to previous conditions. The sensitivity
of the SMT/WFD  was tested with  LED pulses, of  
$\sim6.3$ $\mu$s duration, corresponding to $\beta\sim10^{-4}$,
down to the level of few tens of 
single photoelectrons ({\em spe's}), which is 
the signature of a slow monopole.
A waveform analysis procedure scanned off-line the corresponding wave forms
and simulated in software the function of both the analog
and digital part of the SMT circuitry on an event-by-event basis.
The trigger+analysis efficiency is $> 95\%$ 
for a light yield $> 10$ {\em spe's} over $\sim6.3$ $\mu$s.
 The same waveform methods used to determine
the efficiency of the system are also used to search for
slow monopole signatures in the most recent data.
 Applying simple cuts to reject background from muons
and to guarantee proper functioning of the hardware, we employ
redundant waveform methods to search for slow MMs.
We plan to report on this search in the near future.

\subsubsection{Medium and High Velocity Monopole Searches:}
The data collected by the PHRASE trigger
 are used to search for MMs in the
range $1.2 \cdot 10^{-3} <\beta < 10^{-1}$,  Ambrosio et al. (1992, 1997).
The events
are selected requiring hits in a maximum of four adjacent counters, with a
minimum energy deposition of 10~MeV in two different scintillator layers.
Events with $1.2 \cdot 10^{-3} < \beta < 5 \cdot 10^{-3}$
   are rejected because their
pulse width is smaller than the expected counter crossing time; events
with $5 \cdot 10^{-3} < \beta <
10^{-1}$
 are rejected because the light produced is much lower than 
that expected for a MM.
 The analysis refers to  data 
collected by
the MACRO lower part from October 1989 to the end of 1998 and by the
 Attico from June 1995 to the end of 1998. 
No candidate survives;  the
 $90\%$ C.L. flux upper limit is $3.2 \cdot 10^{-16}$ 
cm$^{-2}$ s$^{-1}$ sr$^{-1}$ (curve ``D" in
Fig. \ref{fig:limiti}a).\par
A previous search for MMs with $\beta >10^{-1}$ based on the ERP trigger, 
Ambrosio et al. (1992,1997)
is  included in Fig. \ref{fig:limiti}a 
( curve `` C''). 

\subsection{Search Using the Streamer Tubes:}
This search was described in 
Ahlen et al. (1995), Ambrosio et al. (1997).  
The analysis is based on the search for single tracks in the streamer
tubes  and on the measurement
of the velocity with the ``time track''. Only the horizontal 
streamer planes of the lower MACRO are used in the 
trigger; the Attico and the  vertical planes 
are used for event  reconstruction.
Data were collected from 1992 to January 1999
for a live-time of 59,712 hours.
The trigger and the analysis chain were checked to be velocity independent. 
The overall efficiency was $\sim$ 73\%. The detector acceptance, computed 
by a Monte Carlo simulation
 including  geometrical and trigger requirements,
is 4250 m$^2$ sr . No monopole candidate was found.
For $1.1 \cdot 10^{-4} < \beta < 5 \cdot 10^{-3}$  the flux upper limit is
$ 3.4 \cdot 10^{-16}$ cm$^{-2}$ s$^{-1}$ sr$^{-1}$
at 90\% C.L.

\subsection{Search Using the Nuclear Track Subdetector:}
The nuclear track subdetector
covers a surface of 1263 m$^{2}$ and the acceptance
for fast MMs is 7100 m$^2$ sr.
 The subdetector is used as a stand alone detector 
and in a ``triggered mode" by 
the scintillator and streamer tube systems. 
The method of searching for MMs and the determination of the
geometrical and detection efficiencies are given in Ahlen et al. (1994).
 An area of 227 m$^2$ of CR39 has been analysed, 
with an average exposure of 7.6 years. 
No candidate was found; the  90$\%$
C.L. upper limits on the MM flux are at the level of 
$6.8 \cdot 10^{-16}$ cm$^{-2}$ s$^{-1}$ sr$^{-1}$ at $\beta\sim$ 1,
and $10^{-15}$ cm$^{-2}$ s$^{-1}$ sr$^{-1}$ at $\beta\sim 10^{-4}$
 (Fig. \ref{fig:limiti}a, curves ``CR39'').

\section{Combined Searches for Fast Monopoles:}

\subsection{} A search for fast MMs with scintillators or streamer tubes
is affected by the background due
to energetic muons with large energy losses. 
A combined use of the three subdetector systems can
achieve the highest rejection by imposing looser requirements.
The trigger requires at least one fired scintillator counter 
and 7 hits in the  horizontal streamer planes.
Candidates are selected on the basis of the scintillator light
yield and  of the digital (tracking) and
analog (pulse charge) information from the streamer tubes.
A further selection is then applied on the streamer tube pulse charge.
After corrections for gain variations,  geometrical 
 and electronic non-linear effects (Battistoni et al.~(1997a)),
a $90\%$ efficiency cut is applied on the average streamer charge. 
 Selected candidates ($\sim$~5/year) are analysed in the
corresponding nuclear track detector modules. The analysis 
refers to about 30,508 live hours 
 with an average efficiency of $77 \%$. 
The geometrical acceptance, computed by Monte Carlo 
methods, including the analysis requirements, is 3565 m$^2$ sr.
No candidate survives; the $90\%$ C.L. flux upper limit is 
 $7.6 \cdot 10^{-16}$ cm$^{-2}$ s$^{-1}$ sr$^{-1}$ for MMs with
$5 \cdot 10^{-3} < \beta < 0.99$ (curve ``E'' in 
Fig. \ref{fig:limiti}a).\par

\subsection{} 
\begin{figure}
   \begin{center}
      \vspace{-2.5cm}
       \mbox{\hspace{-1cm}
              \vspace{-2cm}
              \epsfysize=8cm
               \epsffile{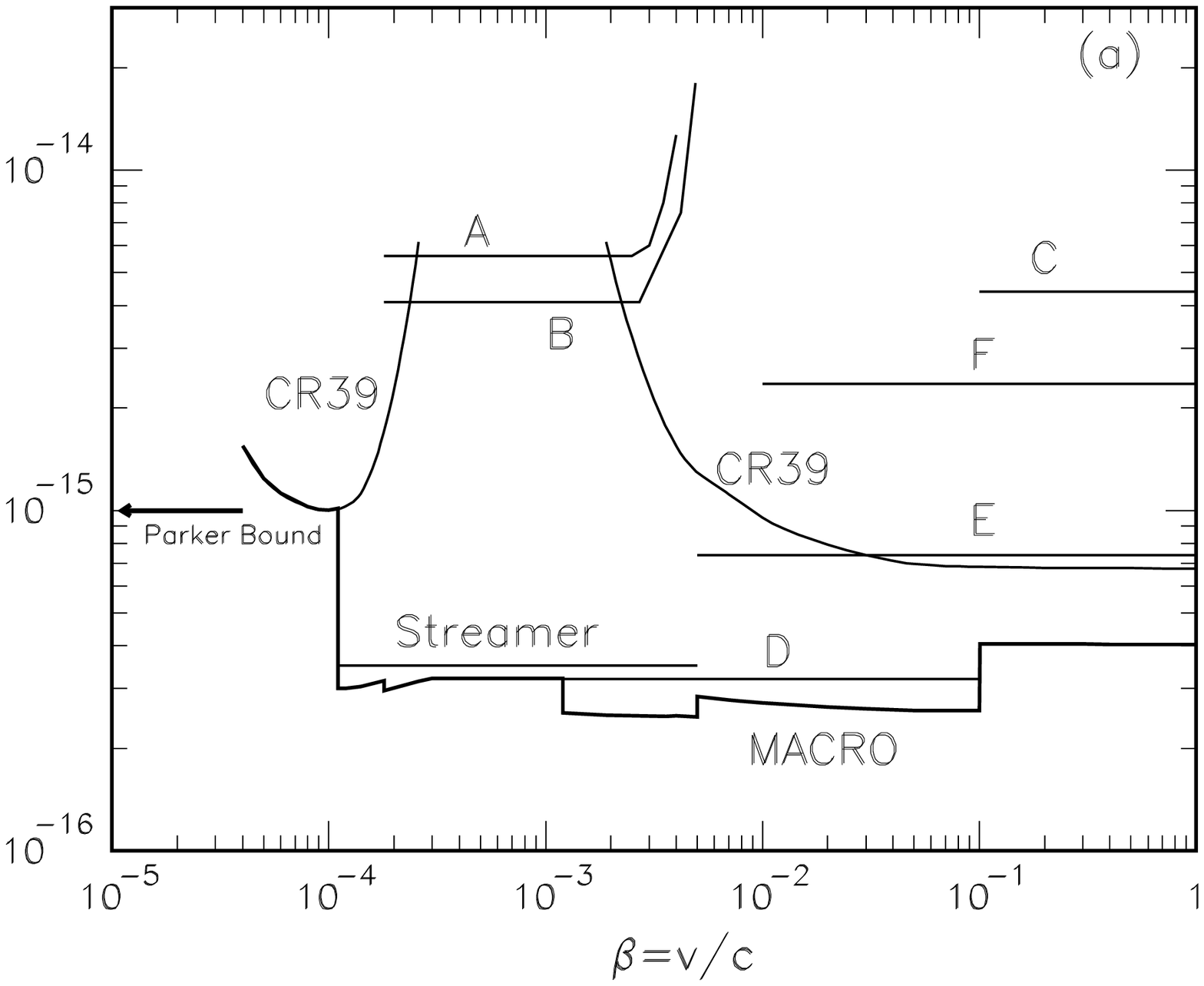}
		\vspace{-2cm}
		\epsfysize=8cm
		\epsffile{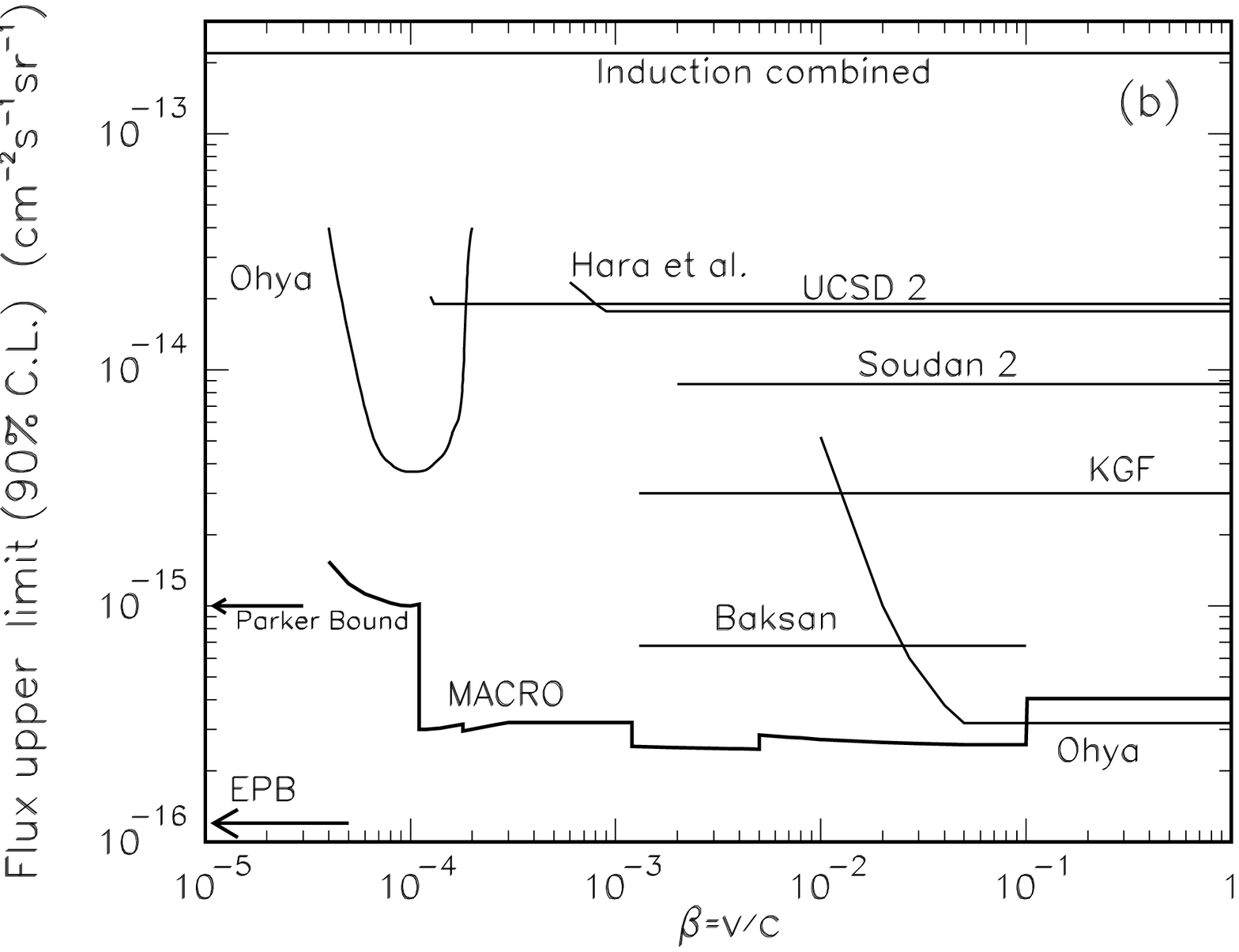}}
   \end{center}
\vspace{-0.5cm}
   \caption{\small The $90\%$ C.L. upper limits obtained (a) using  various
 MACRO subdetectors and the MACRO global limit and (b) by other experiments.}
\label{fig:limiti}
\end{figure}

MMs with $\beta>10^{-2}$ are searched combining 
the streamer tube and  PHRASE triggers. Streamer tubes are 
used to reconstruct the trajectory and pathlength, scintillators are used  
to measure the velocity and the light yield.  Selected events ($\sim 50$/year)
 have
a single track and an energy deposition 
$>$~200~MeV in three scintillator layers.
 The  event energy loss  
is compared to that expected for a 
monopole with the same velocity. The analysis refers to about $8528$ live hours
from May, $1997$ to June, $1998$. No candidate survives.
The geometrical acceptance, including  analysis cuts, is 
3800 m$^2$ sr.  The $90\%$ C.L  flux upper limit is $2.3 \cdot 
10^{-15}$ cm$^{-2}$ s$^{-1}$ sr$^{-1}$ (curve ``F'' in 
Fig. \ref{fig:limiti}a). 

\section{Searches for Nuclearites and Q-balls:}

The searches for MMs based on the
scintillator and nuclear track subdetectors may also be applied
to search for nuclearites, hypothesized nuggets of strange quark matter
and possible candidates for
Dark Matter (DM), Witten (1984).
 Scintillators are sensitive to nuclearites down to
$\beta \simeq 5 \cdot 10^{-5}$ and the CR39
down to $\beta \sim 10^{-5}$,  Ambrosio et al. (1999). 
Individual limits from scintillators and CR39
are presented in 
Fig.~\ref{fig:nucleariti}a (curves ``a-e'' and ``f'', respectively).\par
As recently suggested by Kusenko et al. (1997), the 
MACRO limits for nuclearites may also apply to charged Q-balls (supersymmetric 
coherent states of squarks, sleptons and Higgs fields).
Relic Q-balls are also candidates for cold DM.

\section{Conclusions:}
No MM candidates were found in any search.
 The 90\% C.L. flux limits versus $\beta$ are shown in Fig. \ref{fig:limiti}a. 
 The global MACRO limit is computed as $2.3 / E_{total}$ where 
$E_{total}= \sum_{i}^{} E'_i $,
and the $E'_i$  are the independent time integrated acceptances of 
different analyses. This limit is compared in Fig. \ref{fig:limiti}b with the 
limits of 
other experiments which searched for bare MMs 
 with $g=g_D$ and $\sigma_{cat} <1$ mb, 
 Bermon (1990), Buckland (1990), Thron (1992), Alexeyev (1990), Orito (1991), 
 Adarkar (1990), Hara (1990).\par
\begin{figure}[ht]
   \begin{center}
    \vspace{-1.5cm}
       \mbox{  	\hspace{-1cm}
		\vspace{-1cm}
		\epsfysize=7.4cm
               \epsffile{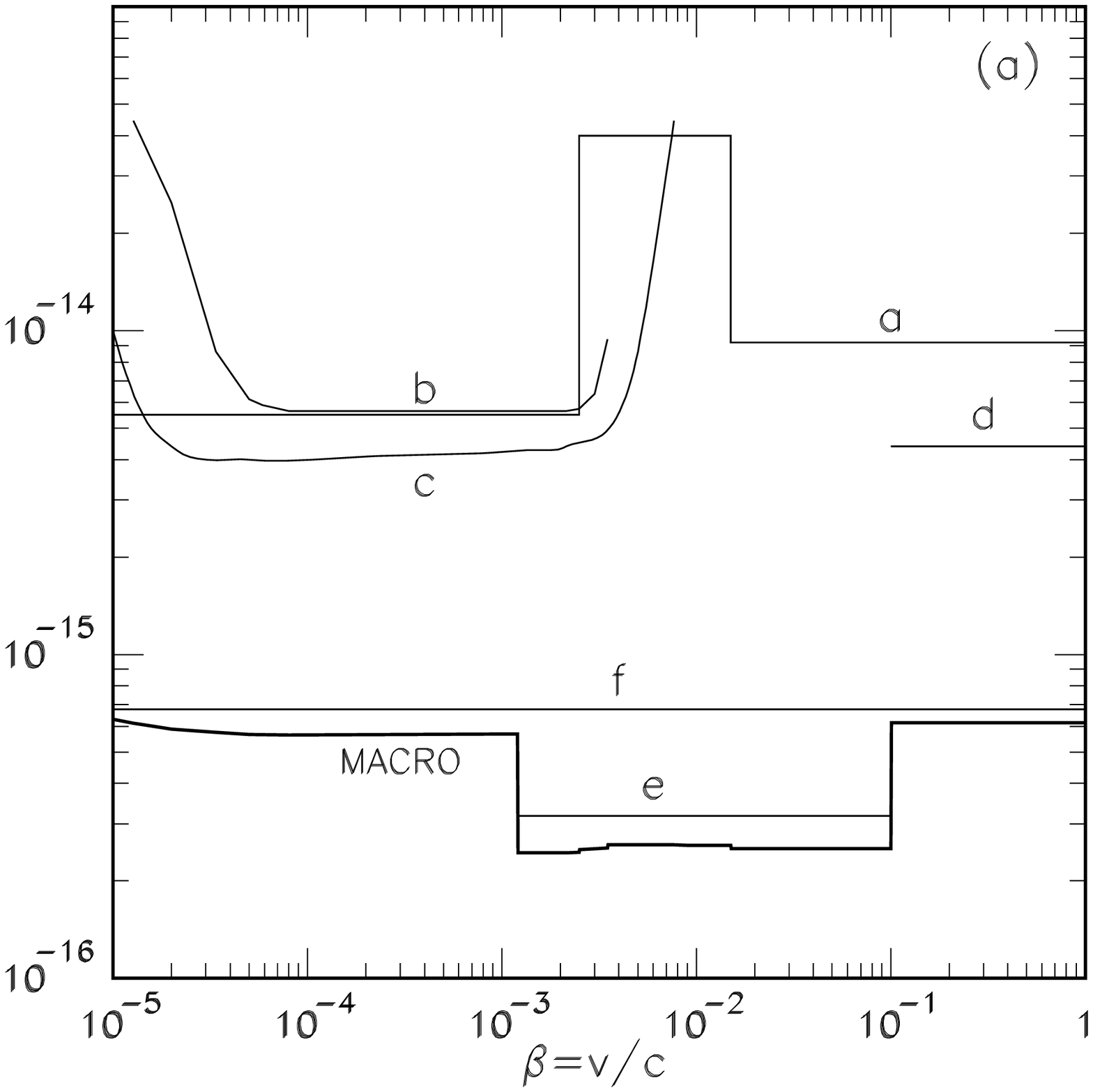}
		\vspace{-1cm}
		\epsfysize=7.5cm
		\epsffile{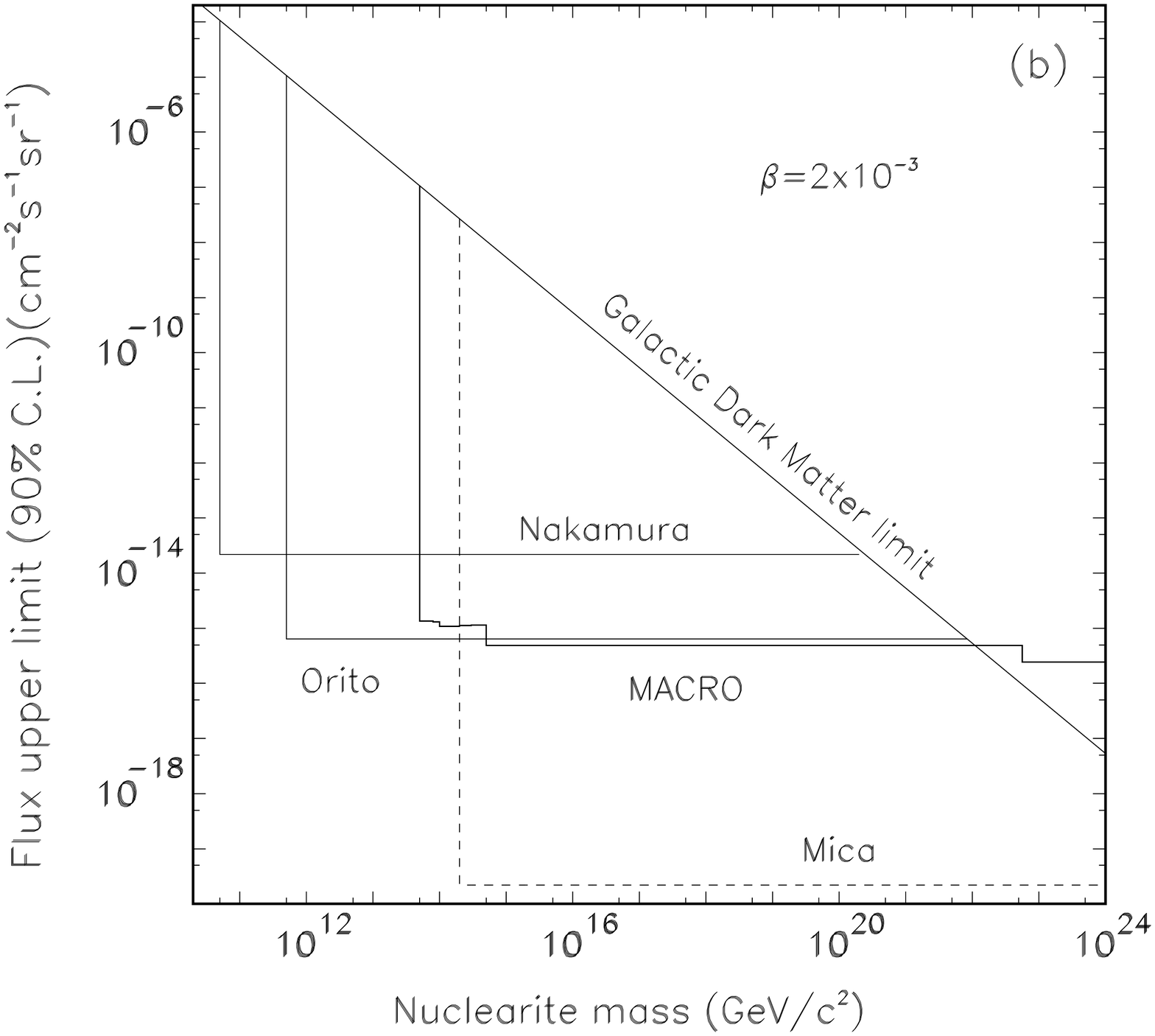}}
   \end{center}
\vspace{-0.5cm}
\caption{\small 90\% C.L. upper limits for an isotropic flux of nuclearites
(a) versus $\beta$ for different MACRO searches with scintillators 
(curves ``a - e'') and CR39 (curve ``f''); (b) 
versus mass obtained by MACRO and by 
other experiments, 
 for downgoing nuclearites with
$\beta = 2 \cdot 10^{-3}$ at ground level. 
The MACRO limit for 
 $M_N > 6 \cdot 10^{22}$ GeV/c$^2$
has been extended above the DM bound and corresponds to an isotropic flux.}
\label{fig:nucleariti}
\end{figure}

 Following the same procedure used for MMs, we obtain the 90\% C.L.
global MACRO limit for an isotropic flux of nuclearites 
(masses  $ > 6 \cdot 10^{22}$ GeV/c$^2$, Fig. \ref{fig:nucleariti}a); at 
$\beta=2\cdot 10^{-3}$ the 
limit is $2.4 \cdot  10^{-16}$ cm$^{-2}$ s$^{-1}$ sr$^{-1}$.
 The MACRO limit for a flux of downgoing
nuclearites is 
compared in Fig.~\ref{fig:nucleariti}b with the limits of other experiments,
Nakamura (1991), Orito (1991), Price (1988), Ghosh \& Chatterjea (1990),
and with the DM bound. 


\newpage
 
%
\vspace{1ex}
\begin{center}
{\Large\bf References}
\end{center}
\vspace{-0.3cm}

\noindent  Adarkar, H. et al., 1990, $21^{st}$ ICRC, Adelaide 95\\
\noindent Ahlen, S. \& Tarl\'e, G. 1983,  Phys. Rev. D27 688\\
\noindent Ahlen, S. P. et al., MACRO Coll., 1993,
 Nucl. Instr. Meth. Phys., A324 337\\
\noindent Ahlen, S. P. et al., MACRO Coll., 1994, LNGS-94/115 \\
 \noindent Ahlen, S. P.  et al., MACRO Coll. 1995, Astrop. Phys. 4 33\\
\noindent Alexeyev, E. N. et al., 1990,  $21^{st}$ ICRC, Adelaide, vol. 10 83\\
\noindent Ambrosio, M. et al.,  MACRO Coll., 1992, Astropart. Phys. 1 11\\
\noindent Ambrosio, M. et al.,  MACRO Coll., 1997, Phys. Lett. B406 \\
\noindent Ambrosio, M. et al.,  MACRO Coll., 1999, hep-ex/9904031, 
EPJ C ( in press)\\
\noindent Battistoni, G. et al., 1988, Nucl. Instr. \& Meth. A270 185\\
\noindent Battistoni, G. et al., 1997a, Nucl. Instr. \& Meth. A401 309\\
\noindent Battistoni, G. et al., 1997b, Nucl. Instr. \& Meth. A399 244\\
\noindent Bermon, S. et al., 1990, Phys. Rev. Lett. 64 839\\
\noindent Buckland, K. N. et al., 1990, Phys. Rev. D41 2726\\
\noindent Cecchini, S. et al., 1996, Nuovo. Cim. A109 1119\\
\noindent Ghosh, D., \ Chatterjea, S., 1990, Europhys. Lett. 12 25\\
\noindent Hara, T. et al., 1990, $21^{st}$ ICRC, Adelaide 79\\ 
\noindent Kusenko, A. et al., 1998, Phys. Rev. Lett. 80 3185\\
\noindent Nakamura, S. et al., 1991, Phys. Lett. B 263 529\\
\noindent Orito, S. et al., 1991, Phys. Rev. Lett. 66 1951\\
\noindent Price, P.~B., 1988, Phys. Rev. D 38 3813\\
\noindent Preskill, J., 1979, Phys. Rev. Lett. 43 1365\\
\noindent Thron, J. L. et al., 1992, Phys. Rev. D46  4846\\
\noindent Turner, M. S., Parker, E. M. \& Bogdan, T. J., 1982,
 Phys. Rev. D26 1926\\
\noindent Witten, E., 1984, Phys. Rev. D30 272\\

\end{document}